\documentclass[12pt]{article}
\usepackage{graphicx}
\usepackage{epsf,amsmath,bbold,amsfonts,stmaryrd}
\usepackage{mathrsfs}
\usepackage{appendix}
\usepackage{amssymb}
\usepackage{float}
\usepackage{color} 
\usepackage[colorlinks]{hyperref}
\hypersetup{pageanchor=false,citecolor=red,urlcolor=red}
\usepackage{indentfirst}
\usepackage[numbers,square,comma,sort&compress,merge]{natbib} 

\usepackage{multibib}
\usepackage[utf8]{inputenc}

\def\a{\alpha}
\def\b{\beta}
\def\g{\gamma}

\def\e{\epsilon}

\def\t{\theta}
\def\m{\mu}
\def\n{\nu}
\def\s{\sigma}

\def\G{\Gamma}

\def\f{\frac}
\def\p{\partial}

\def\l{\left}
\def\r{\right}

\def\Tr{\mathrm{Tr}}

\def\be{\begin{equation}}
\def\ee{\end{equation}}

\def\bea{\begin{eqnarray}}
\def\eea{\end{eqnarray}}

\begin{document}

\begin{titlepage}
\vspace{5cm}

\vspace{2cm}

\begin{center}
\bf \Large{On the strong--CP problem and its axion solution in
torsionful theories}

\end{center}
\vspace{1cm}

\begin{center}
{\textsc {Georgios K. Karananas}}
\end{center}

\begin{center}
{\it Arnold Sommerfeld Center\\
Ludwig-Maximilians-Universit\"at M\"unchen\\
Theresienstra{\ss}e 37, 80333 M\"unchen, Germany}\\

\end{center}

\begin{center}
\texttt{\small georgios.karananas@physik.uni-muenchen.de} \\
\end{center}

\vspace{2cm}

\begin{abstract}

Gravitational effects may interfere with the axion solution to the
strong--CP problem. We point out that gravity can potentially provide
a protection mechanism against itself, in the form of an additional
axion-like field associated with torsion.

\end{abstract}

\end{titlepage}

The effective theory describing the dynamics of the QCD axion $a$
contains a nontrivial interaction between the pseudoscalar and the QCD
Chern-Simons topological density, and has the following schematic
form\,\footnote{Throughout this article we will not keep track of
irrelevant numerical factors.}
\be
\label{eq:eff_action}
\mathscr L_a=\f{1}{2}\partial_\m a \partial^\m a +\f{a}{f} G\widetilde
G \ ,
\ee
with $f$ the axion decay constant. We have used the shorthand notation
\be
G\widetilde G= \f 1 2
\e^{\kappa\lambda\m\n}\Tr\l(G_{\kappa\lambda}{G}_{\m\n}\r) \ ,
\ee
where $\epsilon_{\kappa\lambda\m\n}$ is the totally antisymmetric
symbol, and $G_{\m\n}$ the QCD field strength.

It is clear that~(\ref{eq:eff_action}) is capable of solving the
strong--CP problem, for the minimum of the axion potential forces the
vacuum expectation value of $G\widetilde G$ to vanish. This in turn
implies that physics does not depend on the CP--violating $\t_{\rm
QCD}$ parameter. It should be stressed that this holds true for the
plethora of theories in which this specific coupling of $a$ to the QCD
Chern-Simons term appears, independently of their origin.\footnote{For
example, in the original Peccei--Quinn proposal~\cite{Peccei:1977hh},
the axion emerges as the pseudo--Goldstone boson of a spontaneously
broken anomalous $U(1)$ symmetry.}

It is quite illuminating to show why this is the case by using the
dual formulation of QCD discussed thoroughly in~\cite{Dvali:2005an}
and later in~\cite{Dvali:2013cpa,Dvali:2017mpy}. In this language, the
vacuum superselection problem---or in other words the dependence of
physics on $\t_{\rm QCD}$--- translates into the presence of a
long--range constant field associated with the three-form
\be
\label{eq:qcd_3form}
\mathcal G_{\m\n\lambda}=\Tr\l(A_{[\m}\p_\n
A_{\lambda]}+\f{2}{3}A_{[\m} A_\n A_{\lambda]}\r) \ ,
\ee
where $A_\m$ is the $SU(3)$ gauge field, and the brackets $[\ldots]$
denote antisymmetrization.

In the absence of the axion (as well as massless quarks), the
topological vacuum susceptibility is nonzero~\cite{Kogut:1974kt}
\be
\label{eq:susc_QCD}
\lim_{k\to 0}\int d^4x e^{ikx} \langle E(x)E(0)\rangle\neq 0 \ ,
\ee
where we introduced $E\equiv \e^{\kappa\lambda\m\n}\p_{\kappa}
\mathcal G_{\lambda\m\n}$. At energies below the QCD confinement scale
$\Lambda_{\rm QCD}$, $\mathcal G_{\kappa\lambda\m}$ behaves as a
massless field~\cite{Luscher:1978rn}, since from~(\ref{eq:susc_QCD})
it follows that its propagator has a pole at vanishing virtuality. Its
dynamics is captured by an effective lagrangian, whose (vacuum)
equations of motion dictate that $E=\text{const.}$, in units of
$\Lambda_{\rm QCD}$~\cite{Dvali:2017mpy}. This means that the theory
posseses an infinite number of distinct vacua, one for each value of
$E$.

On the other hand, when the axion is present, then in the dual picture
it is replaced by a two-form $\mathcal A_{\m\n}=-\mathcal A_{\n\m}$,
whose role is to put the massless field~(\ref{eq:qcd_3form}) in a
Higgs phase by providing a (gauge--invariant) mass term for
it.\footnote{In the dual formulation, there is the gauge invariance
\be
\label{eq:gauge_inv_dual}
\mathcal G_{\kappa\lambda\m}\to \mathcal
G_{\kappa\lambda\m}+\p_{[\kappa} c_{\lambda\m]} \ ,~~~\mathcal
A_{\lambda\m}\to \mathcal A_{\lambda\m}+c_{[\lambda\m]} \ ,
\ee
which obviously cannot be broken.} The low-energy dynamics of
$\mathcal G_{\kappa\lambda\m}$ is described
by~\cite{Dvali:2005an,Dvali:2013cpa}
\be
\label{eq:eff_act_dual_QCD}
\mathscr L =\f{E^2}{\Lambda_{QCD}^4} +\f{1}{f^2} (\mathcal
G_{\kappa\lambda\m}-\p_{[\kappa}\mathcal A_{\lambda\m]})^2 \ .
\ee
The fact that the theory has now become gapped, means that the
Chern--Simons field is now screened. This results into the vacuum
suspeptibily being zero, so the physics is independent of $\t_{\rm
QCD}$ and the strong--CP problem is solved.\footnote{This can also be
achieved with massless quarks.}

On general grounds, however, it is believed that gravity violates
global symmetries, the aftermath of which might be the reintroduction
of the strong--CP problem. This can be easily understood, since, in
principle, extra terms---on top of the ones
in~(\ref{eq:eff_action})---can be generated by gravitational
effects. This would result into the axion potential be displaced from
the point where $\langle G\widetilde G\rangle=0$.  In the absence of a
theory of quantum gravity, it seems that there is no way of knowing
the exact form of these contributions.

As realized in~\cite{Dvali:2005an}, the treatment of the problem in
the dual description is particularly suggestive, for it makes clear
that there is a unique way that the axion solution can be
affected. This would correspond to the presence of an additional
three--form field of gravitational origin with a massless pole in its
propagator, which also couples with $\mathcal A_{\m\n}$.

Simply by counting the degrees of freedom in the theory, we notice
that the number of the three--forms in this case would exceed the
number of axions. Thus, necessarily, one of the fields---or better
say, one combination of the fields---will be in a Coulomb phase.

It turns out that the suitable gravitational candidate is the
following three--form
\be
\label{eq:grav_3_form}
\mathcal R_{\m\n\lambda}=
\G^\a_{\b[\m}\p_{\n}\G^\b_{\lambda]\a}+\f{2}{3}\G^\a
_{\b[\m}\G^\b_{\n|\g} \G^\g_{|\lambda]\a} \ ,
\ee
with $\G$ the Christoffel connection. Then, in complete analogy with
QCD, there will be a nonvanishing ``gravitational'' vacuum
susceptibility,
\be
\label{eq:susc_grav}
\lim_{k\to 0} \int d^4 x e^{ikx} \langle E'(x)E'(0)\rangle \neq 0 \ ,
\ee
with $E'\equiv \e^{\kappa\lambda\m\n}\p_\kappa \mathcal
R_{\lambda\m\n}$. The above implies that the vacuum is also permeated
by the constant field $E'\neq 0$; consequently, in the dual picture we
find that the effective theory boils down
to~\cite{Dvali:2005an,Dvali:2013cpa}\,\footnote{Note that, on top
of~(\ref{eq:gauge_inv_dual}), the lagrangian~(\ref{eq:susc_grav}) is
also invariant under the dual version of diffeomorphisms,
\be
\mathcal R_{\kappa\lambda\m} \to \mathcal R_{\kappa\lambda\m} +
\p_{[\kappa}d_{\lambda\m]}\ ,~~~ \mathcal A_{\lambda\m} \to \mathcal
A_{\lambda\m} +d_{[\lambda\m]} \ .
\ee
}
\be
\label{eff_action_dual_QCD_grav}
\mathscr L= \f{E^2}{\Lambda_{QCD}^4} +\f{E'^{\,2}}{\Lambda^4}
+\f{1}{f^2}\l(\a_G \mathcal G_{\m\n\lambda}+\a_R \mathcal
R_{\m\n\lambda}-\p_{[\m}\mathcal A_{\n\lambda]}\r)^2 \ ,
\ee
where $\Lambda$ is a scale set by the correlator~(\ref{eq:susc_grav}),
which need not necessarily be large, and $\a_G$, $\a_R$
constants. \big[As a side note, in the ``conventional picture,'' the
aforementioned mixing between $\mathcal R_{\kappa\lambda\m}$ and
$\mathcal A_{\m\n}$, corresponds to~(\ref{eq:eff_action}) being
supplemented by the term
\be
\f{a}{f}R\widetilde R\ , 
\ee
where
\be
\label{eq:grav_chern-sim_dens}
R\widetilde R = \f 1 2 \e^{\kappa\lambda\m\n}R^\rho_{\
\s\kappa\lambda}{R}^\s_{\ \rho\m\n} \ ,
\ee
is the gravitational parity--odd density. Here, $R^\kappa_{\
\lambda\m\n}$ is the Riemann curvature tensor.\big] 
 We can go a step further and make explicit that the axion solution is
affected. To this end, it is convenient to diagonalize the mass term in
the above by introducing
\be
g_{\m\n\rho}=\a_G \mathcal G_{\m\n\rho}+\a_R \mathcal R_{\m\n\rho} \
,~~~\text{and}~~~ r_{\m\n\rho}=\a_G \mathcal G_{\m\n\rho}-\a_R
\mathcal R_{\m\n\rho}\ .
\ee
Expressed in terms of these new fields, it is easy to see that that
only $g$ gets a mass, while $r$ remains massless. Given the previous
discussion, this is something that should hardly come as a surprise.

A protection mechanism against the gravitational contribution is the
existence of yet another two-form $\mathcal A'_{\m\n}$ in the theory,
such that it screens the second field as
well~\cite{Dvali:2005an,Dvali:2013cpa,Dvali:2017mpy}. For instance,
this can emerge from the presence of neutrinos in the theory, as was
suggested in~\cite{Dvali:2013cpa}. Various aspects of this proposal
were further investigated and generalized
in~\cite{Dvali:2016uhn,*Dvali:2016eay}.

Alternatively, $\mathcal A'_{\m\n}$ can be identified with an
axion-like degree of freedom that couples to $G\widetilde G$, $
R\widetilde R$, or both. It is tempting to entertain the possibility
that this field actually be of gravitational origin. This means that
gravity would have an inherent protection mechanism, which
counterbalances its original effect on the strong--CP problem. Let us
discuss how this can indeed be the case.

It has been known for many years that gauging the Poincar\'e group
yields the Einstein--Cartan--Sciama--Kibble
theory~\cite{Utiyama:1956sy,Brodsky:1962,*Sciama:1962,
*Kibble:1961ba}.\footnote{The Poincar\'e group can be gauged, for
example, by employing the Callan--Coleman--Wess--Zumino coset
construction~\cite{Coleman:1969sm,*Callan:1969sn}, for the case of
spacetime symmetries~\cite{Ivanov:1975zq,*Ivanov:1981wn}. See
also~\cite{Delacretaz:2014oxa,*Karananas:2015eha,*Karananas:2016hrm}
and references therein for a number of generalizations and
applications.} In order to achieve invariance under local translations
and Lorentz transformations, one needs more degrees of freedom than in
General Relativity: the \emph{a priori} independent vielbein and spin
connection, whose respective field stengths are torsion and curvature.

It should be noted that it is in principle possible to eliminate the
extra degrees of freedom by imposing vanishing torsion. In a
four-dimensional spacetime this gives rise to twenty-four constraint
equations that allow to express the connection in terms of the
derivatives of the vielbein (or equivalently the metric).

If, on the other hand, torsion is not eliminated, then the presence of
chiral fermions in this context has quite interesting
implications. The fermionic (torsionful) covariant derivative involves
the axial four-vector of the torsion;\footnote{Torsion can be
decomposed under the Lorentz group into a vector, an axial vector and
a tensor with mixed symmetries.} a rather nontrivial consequence of
this interaction is the emergence of a pseudoscalar axion-like field
$\varphi$, which couples derivatively with the spinorial axial current
$j^\m_5$~\cite{Duncan:1992vz,*Mercuri:2009zi,
*Mercuri:2009zt,*Lattanzi:2009mg,*Castillo-Felisola:2015ema}.\footnote{
Contrary to~\cite{Duncan:1992vz,*Mercuri:2009zi,
*Mercuri:2009zt,*Lattanzi:2009mg,*Castillo-Felisola:2015ema}, however,
\emph{we do not identify} this field with the standard Peccei-Quinn
axion.}

However, it is well known that due to the chiral anomaly, the
divergence of $j^\m_5$ is nonzero.  Consequently, $\varphi$ will
interact (in a clasically shift--symmetric manner) with the
Chern--Simons topological densities associated with QCD and gravity.
It should stressed at this point that the latter mixing appears
obiquitously in the context of torsionful theories, so \emph{its
presence need not be assumed} (for instance
see~\cite{Obukhov:1982da,*Obukhov:1983mm}
and~\cite{Duncan:1992vz,*Mercuri:2009zi,
*Mercuri:2009zt,*Lattanzi:2009mg,*Castillo-Felisola:2015ema}). Let us
note in passing that the divergence of the current might comprise
other terms too, such as the $U(1)$ as well as $SU(2)$ CP--odd
invariants which, nevertheless, are irrelevant for the present
discussion, so we have tacitly ignored them.

In the dual picture, the presence of $\varphi$ with these ``special''
couplings to $G\widetilde G$ and $R\widetilde R$, translates into the
effective theory~(\ref{eff_action_dual_QCD_grav}) becoming
\be
\begin{aligned}
\mathscr L&= \f{E^2}{\Lambda_{QCD}^4} +\f{E'^{\,2}}{\Lambda^4}
+\f{1}{f^2}\l(\a_G \mathcal G_{\m\n\lambda}+\a_R \mathcal
R_{\m\n\lambda}-\p_{[\m}\mathcal A_{\n\lambda]}\r)^2 \\
&+\f{1}{f'^{\,2}}\l(\b_G \mathcal G_{\m\n\lambda}+\b_R \mathcal
R_{\m\n\lambda}-\p_{[\m}\mathcal A'_{\,\n\lambda]}\r)^2 \ .
\end{aligned}
\ee
Here, $\b_G$ and $\b_R$ are constants, while $f'$ is the decay
constant of $\varphi$, which is not a free parameter and its value
is roughly of the order of the Planck
scale~\cite{Duncan:1992vz,*Mercuri:2009zi,
*Mercuri:2009zt,*Lattanzi:2009mg,*Castillo-Felisola:2015ema}.  We
notice that, as long as $\a_G/\a_R\neq\b_G/\b_R$,\footnote{ It would
be somehow peculiar for this relation to not be true in general, since
$\a_G$ and $a_R$ are model--dependent parameters.}  both the QCD as
well as the gravitational three--forms have entered a Higgs phase, so
there are no long--range fields in the vacuum and the solution to the
strong--CP problem persists.

\vspace{1cm}

\begin{center}
\textbf{Acknowledgements}
\end{center}

It is a great pleasure to thank Athanasios Chatzistavrakidis and Gia
Dvali for helpful discussions and important comments on the
manuscript. This work was supported by the ERC-AdG-2013 grant 339169
``Selfcompletion.''  

\newpage

\begin{center}
\textbf{References}
\end{center}

{

\small

\bibliographystyle{utphys}
\bibliography{axi_tors_strong_CP}{}

}

\end{document}